\documentclass[12pt]{iopart}

\usepackage{graphicx}
\usepackage{hyperref}
\usepackage{amssymb}

\newcommand{\im}{\,\mathrm{i}}
\newcommand{\reffig}[1]{figure~\ref{#1}}
\newcommand{\refeq}[1]{equation~(\ref{#1})}

\usepackage{iopams}

\begin{document}

\title[]{Creating electromagnetic cavities using transformation optics}

\author{V~Ginis$^1$, P~Tassin$^{2}$, J~Danckaert$^{1}$, C~M~Soukoulis$^{2,3}$, and I~Veretennicoff$^{1}$}

\address{$^1$ Applied Physics Research Group, Vrije Universiteit Brussel, Pleinlaan~2, B-1050 Brussel, Belgium}
\address{$^2$ Ames Laboratory---U.S. DOE and Department of Physics and Astronomy, Iowa State University, Ames, Iowa 50011, USA}
\address{$^3$ Institute of Electronic Structure and Lasers (IESL), FORTH, 71110 Heraklion, Crete, Greece}

\ead{vincent.ginis@vub.ac.be}

\begin{abstract}
We investigate the potential of transformation optics for the design of novel electromagnetic cavities. First, we determine the dispersion relation of bound modes in a device performing an arbitrary radial coordinate transformation and we discuss a number of such cavity structures. Subsequently, we generalize our study to media that implement azimuthal transformations and we show that such transformations can manipulate the azimuthal mode number. Finally, we discuss how the combination of radial and azimuthal coordinate transformations allows for perfect confinement of subwavelength modes inside a cavity consisting of right-handed materials only.
\end{abstract}


\submitto{\NJP}

\maketitle

\section{Introduction}
The confinement of electromagnetic energy is an essential ingredient in studies of the quantum mechanical properties of light \cite{Walther-2006,Miller-2005,Klaers-2010} as well as in several applications involving the storage and manipulation of information \cite{Matsko-2009,Vahala-2003}. In most circumstances, one requires light to be confined in small volumes over long periods of time. An optical cavity enables confinement of light through internal reflection on its boundaries~\cite{Vahala-2003}. The confinement is, however, only partial because some energy will always be lost to the surrounding environment. Therefore, the eigenmodes of these optical cavities---so-called quasi-normal modes---are characterized by a discrete set of complex eigenfrequencies, where the real part ($\omega^{\prime}$) is proportional to the inverse of the wavelength of the confined light and the imaginary part ($\omega^{\prime \prime}$) is a measure of the temporal confinement of the wave inside the cavity. An important figure of merit is the quality factor $Q$ of these modes, which is usually defined as the temporal confinement of the energy normalized to the frequency of oscillation, such that $Q^{-1}$ represents the fraction of energy lost in a single optical cycle. The quality factor can be calculated as $Q= \omega^{\prime}/(2\omega^{\prime \prime})$. Using specific fabrication techniques, dielectric resonators that exhibit extremely high quality factors have been realized. 
Experimentally, quality factors up to $8\times10^9$ have been measured in dielectric microsphere resonators \cite{Ilchenko-1996} and larger than $10^8$ in dielectric toroid microcavities on a chip~\cite{Armani-2003}. Cavities with high quality factors in combination with small mode volumes are extremely interesting for applications involving cavity quantum electrodynamics~\cite{Hinds-1990,Kimble-1998}, such as the recent developments to integrate optical microresonators into atom chips~\cite{Hinds-2007}. In these applications, it is important to have a small vacuum region in which atoms interact with the electromagnetic modes.
Unfortunately, dielectric cavities are fundamentally limited in size because it is impossible to efficiently store light in volumes with dimensions smaller than the wavelength of the confined mode \cite{Chang-1996, Kavokin-2006}. One attempt to overcome this limitation and to miniaturize the mode volume of the confined light is the development of surface plasmon polariton cavities \cite{Vahala-2009}. Here, however, the temporal confinement is severely reduced by dissipation in the metals. 

In this paper, we show that---using the formalism of transformation optics---alternative designs of subwavelength optical cavities exist. This approach is based on the equivalence between Maxwell's equations in vacuum, expressed in a curved coordinate system, and Maxwell's equations inside a nontrivial material with specific permittivity $\epsilon$ and permeability $\mu$ \cite{Leonhardt-2006, Pendry-2006,Leonhardt-2009}. Although this mathematical equivalence was known for quite some time \cite{Balazs-1957,Plebanski-1960,Felice-1971, Ward-1996}, it was only recently proposed to effectively realize such coordinate transformations with the use of metamaterials \cite{Leonhardt-2006, Pendry-2006}. Transformation optics has demonstrated its huge potential through various proposals of novel optical devices that manipulate the electromagnetic beams in unconventional ways \cite{Chen-2010}. Amongst many others, transformation optics has already been used to design perfect lenses \cite{Shurig-2007,Tsang-2007,Yan-2008,Leonhardt-2010}, beam and polarization manipulators \cite{Rahm3-2008, Kwon-2008}, super scatterers \cite{Wee-2009}, invisibility cloaks \cite{Leonhardt-2006,Pendry-2006, Wei-2008, Leonhardt2-2009, Kildishev-2008}, and devices implementing other optical illusions \cite{Greenleaf-2009,Chan-2009, Lai-2009}. Moreover, the general four-dimensional formulation of transformation optics \cite{Leonhardt2-2006} allows for applications that also involve the time coordinate, such as a frequency convertor \cite{Ginis-2010, Cummer-2010,Miao-2010},  a laser pulse analogue of Hawking radiation \cite{Philbin-2008,Rubino-2011, Faccio-2011}, an electromagnetic analogue of Schwarzschild-(anti-)de Sitter spacetime \cite{Mackay-2011}, or a spacetime cloak \cite{McCall-2011}.

The specific permittivity and permeability tensors required to implement devices designed using the techniques of transformation optics usually do not exist in nature and must, therefore, be achieved with the aid of metamaterials. These man-made artificial materials derive their electromagnetic properties from subwavelength, appropriately designed constituents \cite{Smith-2004}. In particular, metamaterials can be made with a negative pemittivity and permeability at the same frequency. These so-called left-handed materials exhibit peculiar phenomena such as negative phase velocity, negative refraction, and inversed Doppler effect~\cite{Veselago-1968}. Thanks to the ability to compensate the phase of electromagnetic waves inside left-handed materials, they generate the possibility of perfect imaging \cite{Pendry-2000} and miniaturization of optical devices beyond the diffraction limit \cite{Engheta-2002, Alu-2007,Tassin-2008}. More recently, considerable interest has been devoted to metamaterials with other electromagnetic functionalities such as artifical chirality \cite{Soukoulis-2009,Wegener-2009}, a classical analogue of electromagnetically induced transparency \cite{Papasimakis-2008,Tassin-2009,Giessen-2009}, or the enhancement of quantum phenomena \cite{Zheludev-2010}.

In this paper, we will investigate several approaches to design an optical cavity within the framework of transformation optics. In section \ref{Sec:HyperbolicCavity}, we will derive the dispersion relation of a cavity based on a radial coordinate transformation and we will apply this dispersion relation to calculate the bound modes of a cavity in which the radial coordinate is transformed under a hyperbolic function. In section \ref{Sec:PerfectCavity}, we introduce a folded coordinate transformation, which results in a left-handed cavity characterized by a continuum of eigenmodes with an infinite quality factor. Finally, in section \ref{Sec:AzimuthalTransformations}, we introduce a combined transformation on the radial and azimuthal coordinate, which is used in section \ref{Sec:RightPerfectConfinement} to show how perfect confinement can be achieved inside a cavity made of right-handed materials.

\section{A hyperbolic cavity}
\label{Sec:HyperbolicCavity}
\subsection{The dispersion relation of a cylindrical cavity based on a radial coordinate transformation}
\begin{figure}[tbp]
  \centering
   \includegraphics[]{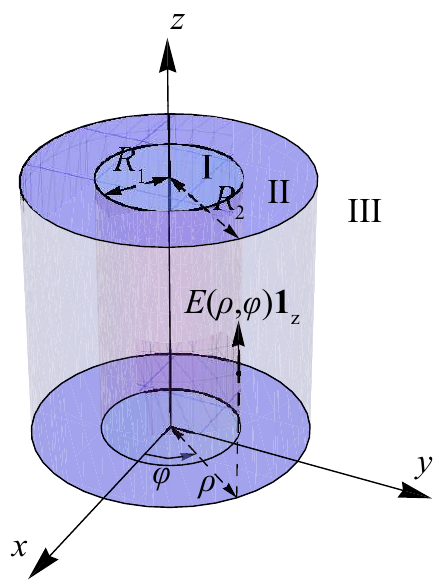} 
  \caption{An infinite hollow cylinder ($\mathrm{II}$) with inner radius $R_1$ and outer radius $R_2$, surrounded by vacuum ($\mathrm{III}$), includes an inner region ($\mathrm{I}$) where electromagnetic fields might be trapped. The medium that performs the coordinate transformation---called ``transformation-optical medium''---is situated in region (II)}
  \label{Fig:CavitySetup}
\end{figure}

We start by calculating the bound modes of the system shown in \reffig{Fig:CavitySetup}, which consists of a hollow cylinder bound by the radii $\rho = R_1$ and $\rho = R_2$. To obtain the eigenmodes of this system, we solve Maxwell's equations inside each region and match the solutions using the proper boundary conditions. Without loss of generality, we will consider the time-harmonic solutions that are polarized along the $z$-axis (TE-polarization): $\mathbf{E}(\mathbf{r},t) = E(\rho,\phi) \exp({-\rmi \omega t}) \,\mathbf{1}_z$.

Inside the empty regions, Maxwell's equations combine into the free-space Helm\-holtz equation for the electric field. The angular variation of the electric field is a sum of imaginary exponentials, characterized by the indices $\nu_\mathrm{I}$ and $\nu_\mathrm{III}$, whereas the radial dependence satisfies the cylindrical Bessel equation. To simplify the boundary conditions in the next step, we will use the Bessel functions $J_{\nu}$ and $Y_{\nu}$ in region (I) and the Hankel functions $H_{\nu}^{(1)}$ and $H_{\nu}^{(2)}$ in the surrounding region (III), because the latter can be interpreted as incoming and outgoing cylindrical solutions of the Bessel equation:
\begin{eqnarray}
\label{Eq:VacuumSolutionI}
E_{\mathrm{I}}(\rho,\phi)=\left[A_{\nu_\mathrm{I}}\ J_{\nu_\mathrm{I}}(k_0\rho)+B_{\nu_\mathrm{I}}\ Y_{\nu_\mathrm{I}}(k_0\rho)\right]\exp(\rmi \nu_\mathrm{I} \phi),\\
\label{Eq:VacuumSolutionIII}
E_{\mathrm{III}}(\rho,\phi)=\left[C_{\nu_\mathrm{III}}\ H_{\nu_\mathrm{III}}^{(1)}(k_0\rho)+D_{\nu_\mathrm{III}}\ H_{\nu_\mathrm{III}}^{(2)}(k_0\rho)\right]\exp(\rmi \nu_\mathrm{III} \phi),
\end{eqnarray}
where $A_{\nu_\mathrm{I}}$, $B_{\nu_\mathrm{I}}$, $C_{\nu_\mathrm{III}}$, and $D_{\nu_\mathrm{III}}$ are complex integration constants and $k_0 = \omega/ c$ represents the vacuum wavenumber.

In order to calculate the solutions in the transformation-optical region (II), we need to insert the constitutive parameters for the material in Maxwell's equations. In transformation optics, these constitutive equations can be derived from the coordinate transformations required to impose a specific pathway onto the electromagnetic fields \cite{Leonhardt-2006,Pendry-2006,Leonhardt-2009}. In this section, we consider the case of an arbitrary radial transformation, leaving the azimuthal angle and the $z$-axis unchanged: ($\rho,\phi,z$) is transformed into ($\rho',\phi',z'$) such that 
\begin{eqnarray}
\rho^{\prime}=f(\rho),\\
\phi^{\prime}=\phi,\\
z^{\prime}=z.
\end{eqnarray}
This distortion of the radial coordinate in vacuum has the same effect on the electromagnetic radiation as if it were propagating in a medium with the following nonzero components of the permittivity and permeability tensors:
\begin{eqnarray}
\epsilon^\rho_{\phantom{\rho}\rho} = \mu^\rho_{\phantom{\rho}\rho}   = \frac{f(\rho)}{\rho f'(\rho)},\nonumber\\
\label{Eq:CylindricalCloakConstitutiveParameters}
\epsilon^\phi_{\phantom{\phi}\phi} = \mu^\phi_{\phantom{\phi}\phi}  = \frac{\rho f'(\rho)}{f(\rho)},\\ 
\epsilon^z_{\phantom{z}z} = \mu^z_{\phantom{z}z} = \frac{f(\rho)f'(\rho)}{\rho},\nonumber 
\end{eqnarray}
where prime denotes differentiation.

Using the constitutive equations $B^i = \mu_0\mu^i_{\phantom{i}j}H^j$ and $D^i = \epsilon_0\epsilon^i_{\phantom{i}j}E^j$, we can insert these parameters in Maxwell's equations and combine them into the following equation for the electric field in region ($\mathrm{II}$):
\begin{equation}
\frac{f(\rho)}{f'(\rho)}\frac{\partial}{\partial \rho}\left(\frac{f(\rho)}{f'(\rho)} \frac{ \partial E_\mathrm{II}}{\partial \rho}\right) + \frac{\partial^2E_\mathrm{II}}{\partial \phi^2} +k_0^2 f^2(\rho)E _\mathrm{II}= 0.
\label{Eq:HelmholtzRegII}
\end{equation}
The solutions of equation (\ref{Eq:HelmholtzRegII}) will of course keep their harmonic azimuthal character. As to the radial part of this equation, it can be reduced to the same Bessel equation as in region (I) and (III), but in the variable $\rho' = f(\rho)$. As a result, the solutions inside the transformation-optical region (II) are given by
\begin{equation}
\label{Eq:EInsideCloak}
E_{\mathrm{II}}(\rho,\phi) = \left[F_{\nu_\mathrm{II}}\ J_{\nu_\mathrm{II}}(k_0f(\rho))+ G_{\nu_\mathrm{II}}\ Y_{\nu_\mathrm{II}}(k_0f(\rho))\right]\exp(\rmi \nu_\mathrm{II} \phi),
\end{equation}
where, once again, $F_{\nu_\mathrm{II}}$ and $G_{\nu_\mathrm{II}}$ are complex integration constants.

This solution can now be matched to the solutions in (I) and (III) [equations (\ref{Eq:VacuumSolutionI}) and (\ref{Eq:VacuumSolutionIII})], using the appropriate boundary conditions. Obviously, both the electric and magnetic fields should be periodic in $\phi$, or $E(\rho,0) = E(\rho,2\pi)$\footnote{This also implies that the magnetic field is periodic in this direction.}. This condition is fulfilled when $\nu_i = m_i$, with $m_i \in \mathbb{Z}$ for all regions $i$. Furthermore, we consider only those modes whose amplitude is finite, which implies that we should reject the Bessel function $Y_\mathrm{m}$ in region (I), because it has a singularity at the origin. In region (III), on the other hand, we impose Sommerfeld's radiation condition, expressing that no energy is flowing in from infinity. We should therefore drop the second Hankel  function $H_m^{(2)}$, which represents such an incoming wave.

The dispersion relation is now found by imposing the continuity of the tangential components of the electric ($E^z$) and magnetic fields ($H^\phi$) at the boundaries between the regions (I), (II), and (III). Firstly, since the boundaries $\rho = R_1$ and $\rho = R_2$ do not depend on $\phi$, the azimuthal mode numbers $m_i$ must be the same in each region. Secondly, we find the following set of four independent equations, in which we already eliminated the angular parts:
\begin{eqnarray}
\label{Eq:DispersionRelationBegin}
A_m\ J_m(k_0R_1) = F_m\ J_m(k_0f(R_1))+G_m\ Y_m(k_0f(R_1)),\\
A_m\ J'_m(k_0R_1) = F_m\ \frac{f(R_1)}{R_1}J'_m(k_0f(R_1))+G_m\ \frac{f(R_1)}{R_1}Y'_m(k_0f(R_1)),\\
F_m\ J_m(k_0f(R_2))+G_m\ Y_m(k_0f(R_2)) = C_m\ H^{(1)}_m(k_0R_2),\label{Eq:DispersionRelation3}\\
\label{Eq:DispersionRelationEnd}
F_m\ \frac{f(R_2)}{R_2}J'_m(k_0f(R_2))+ G_m\ \frac{f(R_2)}{R_2}Y'_m(k_0f(R_2)) = C_m\ H'^{(1)}_m(k_0R_2),
\end{eqnarray}
where $A_{m}$, $F_m$, $G_m$, and $C_{m}$ are complex integration constants.
The equations involving the magnetic field were simplified using the relation $f'(\rho)/\mu^\phi_{\phantom{\phi}\phi} = f(\rho)/\rho$, which is derived from \refeq{Eq:CylindricalCloakConstitutiveParameters}.
Setting the determinant of this set equal to zero generates the dispersion relation of the system and determines the eigenmodes of the cavity. Note that this relation is valid for any cavity of the type shown in figure \ref{Fig:CavitySetup}, implementing a radial coordinate transformation $\rho' = f(\rho)$ between $R_1$ and $R_2$.

\subsection{The confined modes inside a hyperbolic cavity}
A cavity ideally confines the electromagnetic energy in a small (subwavelength) region of space for a very long time. In terms of electromagnetic and physical space, this could be achieved by mapping some large domain of the electromagnetic space onto a much smaller region in physical space. Such a transformation can be constructed with a hyperbolic function that grows to infinity in a finite point. We will therefore consider a device as shown in \reffig{Fig:CavitySetup}, in which the following radial coordinate transformation is implemented between $R_1$ and $R_2$:
$f:\left[R_1,R_2\right] \rightarrow \left[R_1,\infty\right]: \rho \mapsto \rho^\prime$, where
\begin{eqnarray}
f(R_1)=R_1,\label{Eq:HyperbolicCavityFirstBoundary}\\
f(R_2)=\infty.\label{Eq:HyperbolicCavitySecondBoundary}
\end{eqnarray}

\begin{figure}[b]
  \centering
   \includegraphics[]{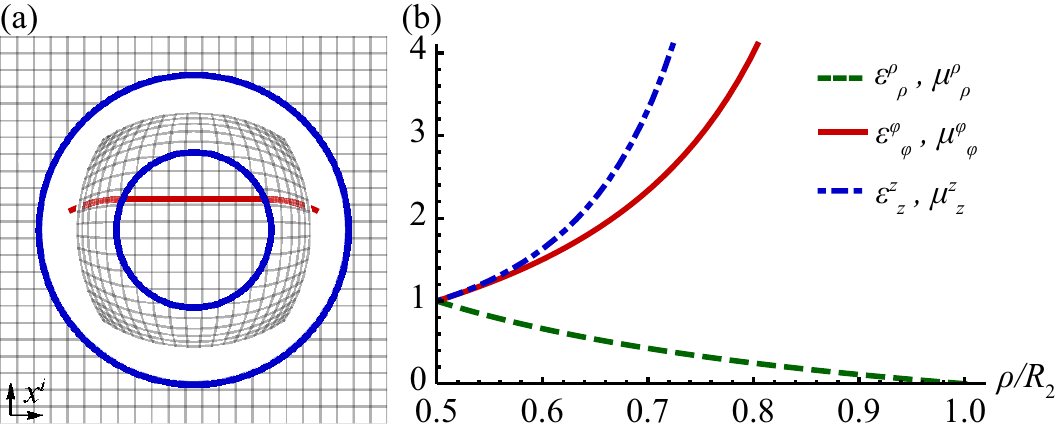}
  \caption{(a) The underlying coordinate transformation of a hyperbolic cavity, defined by $f(\rho)=R_1(R_1-R_2)/(\rho-R_2)$. The region $\left[R_1,R_2\right]$ in the physical radial coordinate $\rho$ covers the region $\left[R_1, +\infty \right]$ in the electromagnetic radial coordinate $\rho^\prime$. (b) The material parameters that implement this hyperbolic transformation. In the limiting case of $f(R_2) \rightarrow \infty$ the radial components of $\epsilon$ and $\mu$ become zero, while the other parameters grow to infinity at the outer boundary.}
  \label{Fig:HyperbolicMapTransformationFunction}
\end{figure}

The matching at the inner boundary enables a smooth transition of the waves. Since there cannot be anything ``beyond infinity,'' equation (\ref{Eq:HyperbolicCavitySecondBoundary}) ensures that the electromagnetic energy cannot escape this device. Such a coordinate transformation is illustrated in figure \ref{Fig:HyperbolicMapTransformationFunction}(a), where we consider the transformation given by $f(\rho) = R_1(R_1-R_2)/(\rho-R_2)$. The Cartesian coordinate lines in physical space become denser as we approach the outer radius $R_2$. A similar transformation has been proposed to design a matching layer in order to improve the efficiency of numerical software algorithms \cite{Zharova-2008}. The values of the material parameters required to implement this transformation in physical space are shown in figure \ref{Fig:HyperbolicMapTransformationFunction}(b).

The determinant of equations (\ref{Eq:DispersionRelationBegin})-(\ref{Eq:DispersionRelationEnd}) can be calculated in the limit of $f(R_2) \rightarrow \infty$, using equation (\ref{Eq:HyperbolicCavityFirstBoundary}). We find that the hyperbolic map does not confine any electromagnetic modes. In contrast to what is mentioned in reference~\cite{Zhai-2010}, we find that, independent of the azimuthal mode number $m$, the dispersion relation can only be satisfied if $k_0 = 0$, i.e., the static solution. This result fits with the intuitive idea that in this configuration an electromagnetic wave travels an infinitely long time to reach the outer boundary of the cavity. Therefore, no standing wave can be created in the cavity: its structure does not permit a reflected wave at $\rho = R_2$.

\begin{figure}[tb]
  \centering
   \includegraphics[]{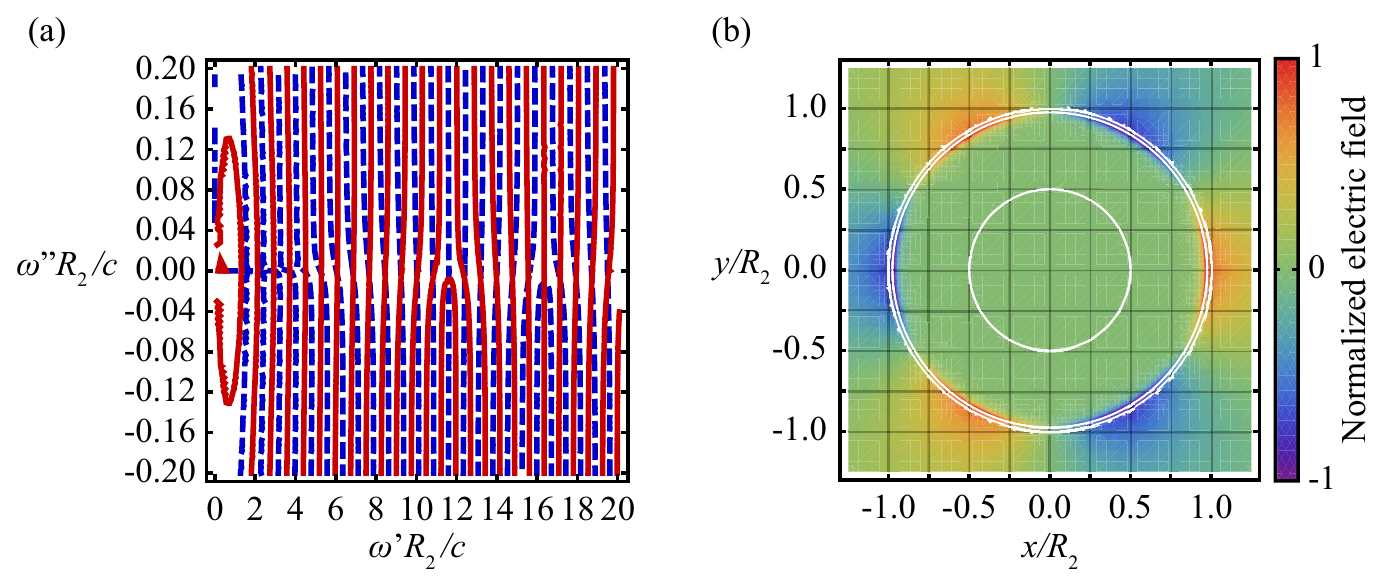} 
  \caption{The confined modes with azimuthal mode number $m=3$ of a perturbed hyperbolic map, in which we cut out a rim $\Delta R$ of the outer boundary $R_2$, such that the outer value $R_\mathrm{out}=R_2-\Delta R$ is not mapped onto infinity, but instead takes the value $f(R_2-\Delta R)=20$. (a) A contour plot in de complex frequency plane indicating the solutions of the dispersion relation. (b) The two-dimensional plot of the electric field corresponding to the first solution of the dispersion relation.}
  \label{Fig:NonIdealHyperbolicMap}
\end{figure}

The previous physical interpretation implies that a perturbed version of the hyperbolic map---in which the material parameters do not grow to infinity---should exhibit confined modes. This is indeed confirmed by the numerical evaluation of the dispersion relation, whose solutions are shown in figure \ref{Fig:NonIdealHyperbolicMap}. The number of solutions increases as $\Delta R$ decreases, or equivalently as $f(R_2-\Delta R)$ approaches $\infty$. The quality factor $Q$ increases at the same pace. As shown in figure \ref{Fig:NonIdealHyperbolicMap}, the cavity enables subwavelength confinement of electromagnetic energy; the first solution, for example, lies at $k_0 R_2 = 0.51 - 7.0\times10^{-6}\, \rmi$, which corresponds to a free space wavelength $\lambda_0$ that is more than $10$ times larger than the outer radius of the cavity $R_2$. The quality factor of this mode is $Q = 3.6\times10^{4}$.

In figure \ref{Fig:NonIdealHyperbolicMap}(b), we show a two-dimensional plot of this mode inside the cavity. The reader should note how the field is almost completely located in the transformation-optical medium, which sounds reasonable since this medium contains the electromagnetic interval $\left[R_1,+\infty \right]$, whereas the inner disk (region $\mathrm{I}$) only occupies the electromagnetic interval $\left[0,R_1\right]$; inside the transformation-optical region the wavelength of the electric field becomes smaller towards the outer radius $R_2$ due to the increasing values of the material parameters inside the medium.

Judging from these results, the imperfect hyperbolic design might seem to be the subwavelength optical cavity we are looking for, having good confinement in arbitrarily small dimensions. We should, however, look back at the materials with which it is implemented in figure \ref{Fig:HyperbolicMapTransformationFunction}(b). Firstly, we note that a traditional whispering gallery cavity, made entirely from these high-index materials, also has subwavelength modes. Secondly, the wavelength is becoming extremely small within the device so that it is practically impossible to use the mean-field approximation when determining the material parameters. In the subsequent sections, we will present subwavelength cavities in which this is no longer the case.

\section{The perfect cavity}
\label{Sec:PerfectCavity}
\begin{figure}[t]
  \centering
   \includegraphics[]{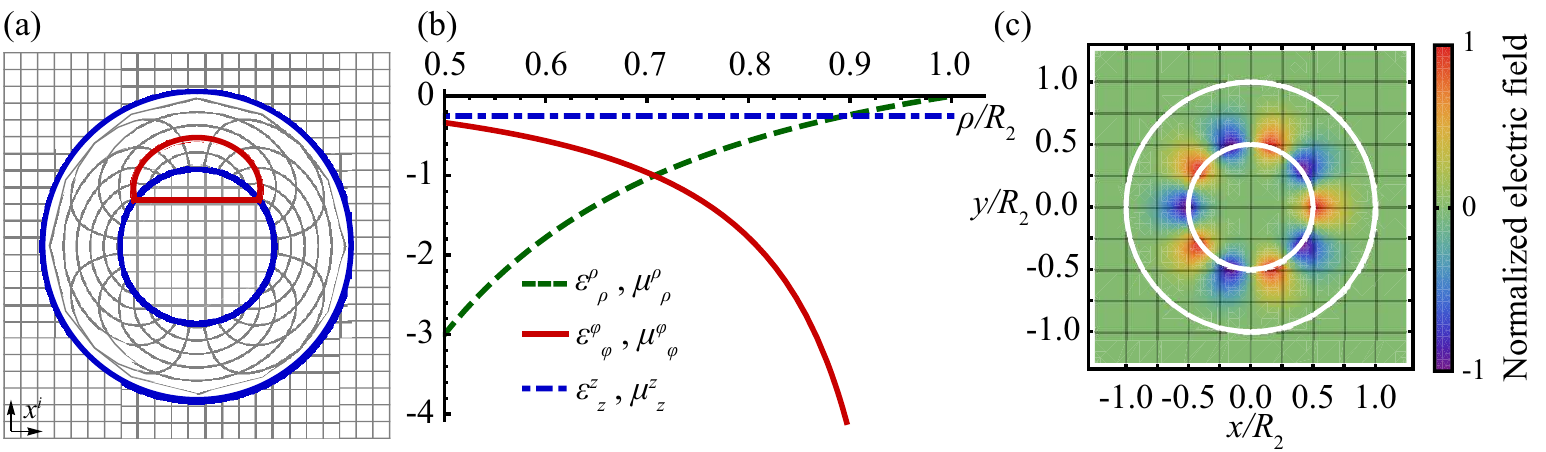} 
  \caption{(a) The coordinate lines that are generated by a radial coordinate transformation implementing a perfect cavity, defined by $f(\rho) = \frac{R_1}{R_1-R_2}(\rho-R_2)$. (b)~The material parameters required to materialize this coordinate transformation. (c)~The electric field mode profile of a confined mode inside the cavity. The wavelength of this mode is much larger (factor 100) than the outer radius of the cavity.}
    \label{Fig:PerfectCavity_Solutions}
\end{figure}

In the example of the hyperbolic cavity, the entire electromagnetic space was mapped onto a finite region of physical space. We can, however, consider a cavity from a cloaking perspective and design a device that cloaks away the volume surrounding the device, instead of the volume inside the device \cite{Ginis-2010-2}. Such a device should smoothly guide the electromagnetic waves in the cavity so that they never penetrate the outer boundary. 
The effect of such a transformation is shown in figure \ref{Fig:PerfectCavity_Solutions}(a). Since we want to cloak  away region~$(\mathrm{III})$, we will use a radial coordinate transformation that maps the physical coordinates $(\rho,\phi,z)$  onto the electromagnetic coordinates $(\rho',\phi',z')$. To achieve perfect cloaking of region $(\mathrm{III})$ from the viewpoint of region $(\mathrm{I})$, the radial transformation function has to satisfy the following boundary requirements:
\begin{eqnarray}
\label{Eq:TransformationFunctionBoundary1}
f(R_1) & = R_1,\\
\label{Eq:TransformationFunctionBoundary2}
f(R_2) & = 0.
\end{eqnarray}
As before, the actual shape of the function has no implications on the cavity's performance. Transformation functions that satisfy these boundary conditions have also been studied in combination with traditional invisibility cloaks, giving rise to so-called anti-cloaks \cite{Chan-2008, Engheta-2009}.

The modes of the present cavity are the solutions of equations~(\ref{Eq:DispersionRelationBegin})-(\ref{Eq:DispersionRelationEnd}), where we now have to insert $f(R_1) = R_1$ and $f(R_2) = 0$. These equations now become
\begin{eqnarray}
\label{Eq:PerfectCavityDispersionRelationBegin}
& A_m\ J_m(k_0R_1) = F_m\ J_m(k_0 R_1)+G_m\ Y_m(k_0 R_1),\label{Eq:PerfCafIntr1}\\
& A_m\ J'_m(k_0R_1) = F_m\ J'_m(k_0 R_1)+G_m\ Y'_m(k_0 R_1),\label{Eq:PerfCafIntr2}\\
& F_m\ \lim_{x \to 0}J_m(k_0 x)+G_m\ \lim_{x \to 0}Y_m(k_0 x) = C_m\ H^{(1)}_m(k_0R_2),\label{Eq:PerfCafIntr3}\\
\label{Eq:PerfectCavityDispersionRelationEnd}
& F_m\ \lim_{x \to 0}\left[\frac{x}{R_2}J'_m(k_0 x)\right]+ G_m\ \lim_{x \to 0}\left[\frac{x}{R_2}Y'_m(k_0 x)\right] = C_m\ H'^{(1)}_m(k_0R_2).\label{Eq:PerfCafIntr4}
\end{eqnarray}
These limits should be handled with care, since they contain indefinite expressions like $0 \times \infty$. Assuming the azimuthal mode number $m \neq 0$, these limits can be unambiguously evaluated:
\begin{eqnarray}
&\lim_{x \to 0}J_m(k_0 x) = 0,\\
&\lim_{x \to 0}Y_m(k_0 x) = -\infty,\\
&\lim_{x \to 0}\left[\frac{x}{R_2}J'_m(k_0 x)\right] = 0,\\
&\lim_{x \to 0}\left[\frac{x}{R_2}Y'_m(k_0 x)\right] = +\infty.
\end{eqnarray}
We can now reinsert these limits in equations (\ref{Eq:PerfCafIntr3})-(\ref{Eq:PerfCafIntr4}) and we find that this set only has solutions if $G_m = 0$ and $C_m = 0$ for all azimuthal mode numbers $m\neq0$, whereas there are no requirements on $F_m$. Taking this into account, equations (\ref{Eq:PerfCafIntr1}) and (\ref{Eq:PerfCafIntr2}) become
\begin{eqnarray}
\label{Eq:PerfectCavityTrivialSetBegin}
& A_m\ J_m(k_0R_1) = F_m\ J_m(k_0 R_1),\\
\label{Eq:PerfectCavityTrivialSetEnd}
& A_m\ J'_m(k_0R_1) = F_m\ J'_m(k_0 R_1),
\end{eqnarray}
hence $A_m = F_m.$
This set imposes no constraints on $k_0$, which means that the cavity supports a continuous spectrum of modes, even if the wavelength is larger than the characteristic dimensions of the cavity. These modes are perfectly confined, since $D$ is equal to zero: there is no radiation escaping into region (III). The quality factor $Q$ is infinite and, as a consequence, the complex part of the frequency ($\omega"$) should be zero.

We are now able to plot the solutions of the perfect cavity. We can choose any real free-space wave vector $k_0$ and plot the solutions, using \refeq{Eq:EInsideCloak}. In figure \ref{Fig:PerfectCavity_Solutions}(c), we plot a mode for which $k_0R_1 = 0.01$. The field's variation inside the cavity depends on the chosen transformation function $f(\rho)$. One can make well-considered choices for this function $f$ to enhance the field distribution inside the transformation-optical medium.

We observe a completely different mechanism of confinement as compared to the hyperbolic map. Generally, a wave can be confined inside a cavity if one round trip (approximately the cavity's circumference) equals an integer number $l$ of the mode's wavelength inside the cavity: $2\pi a \approx  l \lambda$ \cite{Chang-1996}. The perturbed version of the hyperbolic map reduces the wavelength of an electromagnetic mode to a very small number at the outer boundary, thus fulfilling the condition. In the perfect cavity, however, the phase shift vanishes completely and $l=0$.

The reduction of equations (\ref{Eq:PerfCafIntr1})-(\ref{Eq:PerfCafIntr4}) to the trivial equations (\ref{Eq:PerfectCavityTrivialSetBegin})-(\ref{Eq:PerfectCavityTrivialSetEnd}) was only possible when we assumed the azimuthal mode number $m \neq 0$. A mode without azimuthal momentum cannot be confined within this cavity. Physically, this can be understood since such a mode has a purely radial wave vector and in the absence of azimuthal propagation it cannot be deflected to the left or to the right inside the transformation-optical region.

The material losses are high due to the fact that the transformation-optical medium is made of left-handed materials, as shown in figure \ref{Fig:PerfectCavity_Solutions}(b). Although the material parameters strongly depend on the choice of the transformation function $f(\rho)$---through equations~(\ref{Eq:CylindricalCloakConstitutiveParameters})---one can prove that any transformation-optical medium that satisfies equations~(\ref{Eq:TransformationFunctionBoundary1})-(\ref{Eq:TransformationFunctionBoundary2}) will have a region in which all components of the permitivity and the permeability tensors are negative. This is analogous to the perfect lens \cite{Pendry-2000, Tassin-2006}---another example of a folded map \cite{Leonhardt-2009}---which also requires a left-handed response.

In the last sections of this article, we will derive a method to overcome this limitation and demonstrate how it is possible to design a cavity with right-handed material parameters only. But let us first introduce the idea of cavities based on azimuthal coordinate transformations.

\section{Azimuthal coordinate transformations}
\label{Sec:AzimuthalTransformations}
\subsection{The dispersion relation in case of an azimuthal coordinate transformation}
In this section, we investigate transformation-optical cavities, as shown figure \ref{Fig:CavitySetup}, in which the transformation also involves the azimuthal coordinate $\phi$. We will consider a transformation defined by
\begin{eqnarray}
\rho^{\prime}=f(\rho),\\
\phi^{\prime}=g(\phi),\\
z^{\prime}=z.
\end{eqnarray}
and, once again, look at solutions of Helmholtz' equation with linear polarization along the $z$-axis.
It can be shown that such a transformation can be implemented with materials whose components are
\begin{eqnarray}
\epsilon^\rho_{\phantom{\rho}\rho} = \mu^\rho_{\phantom{\rho}\rho}   = \frac{f(\rho)}{\rho f^{\prime}(\rho)}g^{\prime}(\phi),\nonumber\\
\epsilon^\phi_{\phantom{\phi}\phi} = \mu^\phi_{\phantom{\phi}\phi}  = \frac{\rho f^{\prime}(\rho)}{f(\rho)}\frac{1}{g^{\prime}(\phi)},\\ 
\epsilon^z_{\phantom{z}z} = \mu^z_{\phantom{z}z} = \frac{f(\rho)f^{\prime}(\rho)}{\rho}g^{\prime}(\phi),\nonumber 
\end{eqnarray}
in which $f^{\prime}(\rho)$ denotes differentiation of $f(\rho)$ with respect to $\rho$ and $g^{\prime}(\phi)$ denotes differentiation of $g(\phi)$ with respect to $\phi$. 
The wave equation of such a medium is
\begin{equation}
\frac{f(\rho)}{f'(\rho)}\frac{\partial}{\partial \rho}\left(\frac{f(\rho)}{f'(\rho)} \frac{ \partial E}{\partial \rho}\right) + \frac{1}{g^{\prime}(\phi)}\frac{\partial}{\partial \phi}\left(\frac{1}{g^{\prime}(\phi)}\frac{\partial E}{\partial \phi}\right) +k_0^2 f^2(\rho)E = 0,
\end{equation}
whose solutions are given by
\begin{equation}
\label{Eq:EInsideCloak2}
E_{\mathrm{II}}(\rho,\phi) = \left[F_\nu\ J_{\nu_{\mathrm{II}}}(k_0f(\rho))+ G_\nu\ Y_{\nu_{\mathrm{II}}}(k_0f(\rho))\right]\exp(\rmi \nu_\mathrm{II} g(\phi)).
\end{equation}
Here again, the cylindrical symmetry leads to the quantization of the azimuthal mode number $\nu_{\mathrm{II}} = m_{\mathrm{II}}$:
\begin{equation}
\label{Eq:AngularModeNumberQuantization}
m_{\mathrm{II}}(k) = \frac{2 \pi k}{g(2\pi)-g(0)},
\end{equation}
with $k \in \mathbb{Z}$. However, unlike the dispersion relation derived in previous sections where the mode numbers in the different regions were identical, a general azimuthal transformation will scramble the azimuthal momenta of the solutions in the different regions. One single azimuthal mode $\exp(\im m_\mathrm{II} g(\phi))$ in the transformation-optical region (II) will excite multiple modes in the vacuum region, and vice versa:
\begin{equation}
\label{Eq:FourierSeriesExpansion}
\exp(\im m_\mathrm{II} g(\phi)) = \sum_{m_\mathrm{I}} C_{m_\mathrm{I}} \exp(\im m_\mathrm{I}\phi),
\end{equation}
where the coefficients $C_{m_\mathrm{I}}$ are given by
\begin{equation}
\label{Eq:ModenumberScramble}
C_{m_\mathrm{I}} = \frac{1}{2\pi}\int_0^{2\pi}\exp(\im \left(m_\mathrm{II}g(\phi)-m_\mathrm{I}\phi\right))\mathrm{d}\phi.
\end{equation}
The Fourier series expansion in equations (\ref{Eq:FourierSeriesExpansion})-(\ref{Eq:ModenumberScramble}) is possible since $\exp(\im m_\mathrm{II} g(\phi))$ is a periodic function of the azimuthal coordinate $\phi$ with period $2\pi$, as can be seen by substituting equation (\ref{Eq:AngularModeNumberQuantization}) into equation (\ref{Eq:EInsideCloak2}).
In the surrounding vacuum region (III), the same condition on the azimuthal coordinate applies, $C_{m_\mathrm{I}}=C_{m_\mathrm{III}}$.

We will calculate now the dispersion relation and restrict the analysis to linear azimuthal transformations $g(\phi) = a\phi$, where $a$ is a real number. Using equation (\ref{Eq:ModenumberScramble}), it can be shown that in this case a single azimuthal mode number $m_\mathrm{I}=m_\mathrm{III}=m_1$ in the vacuum regions will match with a single mode number $m_\mathrm{II}=m_2$ in the transformation-optical region, where these mode numbers are related by
\begin{equation}
m_2 = \frac{m_1}{a}.
\end{equation}
In general, the angular mode number $m_2$  will not be an integral number.
The dispersion relation is then similar to the one that corresponds to a single radial coordinate transformation (\ref{Eq:DispersionRelationBegin})-(\ref{Eq:DispersionRelationEnd}) and is generated by the following set of equations:
\begin{eqnarray}
\label{Eq:DispersionRelationRadialAngularBegin}
A_{m_1}\ J_{m_1}(k_0R_1) = F_{m_2}\ J_{m_2}(k_0f(R_1))+G_{m_2}\ Y_{m_2}(k_0f(R_1)),\\
\label{Eq:DispersionRelationRadialAngularMiddle1}
A_{m_1}\ J'_{m_1}(k_0R_1) = F_{m_2}\ \frac{f(R_1) a}{R_1}J'_{m_2}(k_0f(R_1))\nonumber\\ \qquad\qquad\qquad\qquad+G_{m_2}\ \frac{f(R_1) a}{R_1}Y'_{m_2}(k_0f(R_1)),\\
\label{Eq:DispersionRelationRadialAngularMiddle2}
F_{m_2}\ J_{m_2}(k_0f(R_2))+G_{m_2}\ Y_{m_2}(k_0f(R_2)) = C_{m_1}\ H^{(1)}_{m_1}(k_0R_2),\\
\label{Eq:DispersionRelationRadialAngularEnd}
F_{m_2}\ \frac{f(R_2) a}{R_2}J'_{m_2}(k_0f(R_2))+ G_{m_2}\ \frac{f(R_2) a}{R_2}Y'_{m_2}(k_0f(R_2))\nonumber\\ \qquad\qquad\qquad\qquad= C_{m_1}\ H'^{(1)}_{m_1}(k_0R_2).
\end{eqnarray}
The additional factors $a =g^{\prime}(\phi)$ in equations (\ref{Eq:DispersionRelationRadialAngularMiddle1}) and (\ref{Eq:DispersionRelationRadialAngularEnd}) originate from the fact that
\begin{equation}
\frac{f^{\prime}(\rho)}{\mu^{\phi}_{\phantom{\phi}\phi}} =  \frac{f^{\prime}(\rho)f(\rho)}{\rho f^{\prime}(\rho)}g^{\prime}(\phi)= \frac{f(\rho)a}{\rho}. 
\end{equation}

\subsection{Cavities based on a single azimuthal transformation}
\begin{figure}[t]
  \centering
   \includegraphics[]{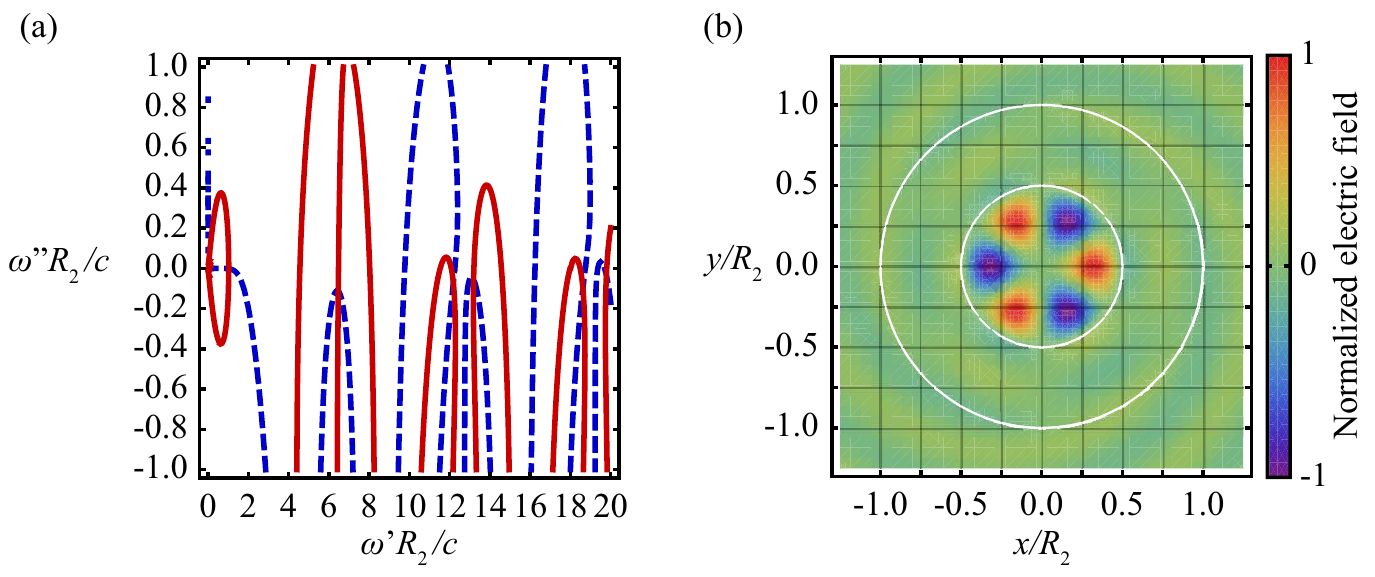} 
  \caption{(a) Contour plot of the dispersion relation of the confined modes whose azimuthal mode number in region (I) equals $m_1=5$. The cavity is defined by equation (\ref{Eq:AngularCavity_Example}). (b) A density plot of the electric field distribution inside the cavity, corresponding to the fourth solution in (a), at $k_0 = 13 - 5.7 \times10^{-2} \im$.}
    \label{Fig:AngularCavity_Solutions}
\end{figure}

It is instructive to have a look at the modes of a cavity defined by an exclusively azimuthal transformation, for instance,
\begin{eqnarray}
\label{Eq:AngularCavity_Example}
g(\phi)=15\phi,\\
f(r) = r.
\end{eqnarray}
As shown in figure \ref{Fig:AngularCavity_Solutions}(a), there are several confined modes within these cavities.  A corresponding mode profile is shown in figure \ref{Fig:AngularCavity_Solutions}(b). Very much in correspondence with the hyperbolic map on the radial coordinate in section \ref{Sec:HyperbolicCavity}, we find that the quality factor of the solutions increases as the optical path length inside the cavity increases. In contrast to the hyperbolic map, however, the number of subwavelength solutions does not drastically increase as we increase the optical path length. 

Another intriguing example is the collapse of the azimuthal coordinate, i.e., all angles are transformed on one and the same angle ($g(\phi) = \phi_0$), inside a full cylinder (no vacuum region). Although the corresponding material parameters are extremely exotic (zero and infinity) and thus not useful for practical applications, this setup is interesting for theoretical reasons. Obviously, there will be no quantization of the azimuthal mode number~$m$, and the continuity relations imply that
\begin{eqnarray}
& \int_{-\infty}^{+\infty} F(m)\ J_{m}(k_0 R)\mathrm{d}m  = C\ H^{(1)}_0(k_0R),\\
& 0 \times \int_{-\infty}^{+\infty} F(m) \ J^{\prime}_{m}(k_0 R)\mathrm{d}m = C\ H'^{(1)}_0(k_0R).
\end{eqnarray}
which immediately translates into
\begin{equation}
\int_{-\infty}^{+\infty} F(m)\ J_{m}(k_0 R)\mathrm{d}m  = 0.
\end{equation}
This equation has a solution for any $k_0$. The cavity thus confines light at every wavelength. These two examples clearly show the difference between a radial and an azimuthal transformation. The former changes the radial coordinate $\rho $, which automatically alters the quantization of $k_0$, whereas the latter manipulates $\phi$, which changes the azimuthal mode number $m$ and thus only indirectly influences the quantization of $k_0$ through the dispersion relation.

\section{Perfect confinement in a right-handed cavity}
\label{Sec:RightPerfectConfinement}
Azimuthal transformation optics can be very valuable when used in combination with a nontrivial radial transformation. To demonstrate this, we show here how the addition of an azimuthal transformation can be used to generate a cavity with right-handed material parameters in which there are no radiation losses. Let us consider the radial  transformation of a perfect cavity, in combination with an azimuthal transformation that inverts $\phi$:
\begin{eqnarray}
f(\rho) = \frac{R_1}{\sqrt{R_1^2-R_2^2}}\sqrt{\rho^2-R_2^2},\\
g(\phi) = -\phi,
\end{eqnarray}
implemented between the radii $\rho = R_1$ and $\rho = R_2$.

The material parameters will be the same as those of a perfect cavity. However, due to the inversion of $\phi$ an additional sign reversal will make all material parameters positive. The dispersion relation of this cavity is then given by the following set of equations:
\begin{eqnarray}
\label{Eq:PositivePerfectCavityDispersionRelationBegin}
A_m\ J_m(k_0R_1) = F_{-m}\ J_{-m}(k_0 R_1)+G_{-m}\ Y_{-m}(k_0 R_1),\\
\label{Eq:PositivePerfectCavityDispersionRelationMiddle1}
A_m\ J'_m(k_0R_1) = F_{-m}\ (-1)J'_{-m}(k_0 R_1)+G_{-m}\ (-1)Y'_{-m}(k_0 R_1),\\
\label{Eq:PositivePerfectCavityDispersionRelationMiddle2}
F_{-m}\ \lim_{x \to 0}J_{-m}(k_0 x)+G_{-m}\ \lim_{x \to 0}Y_{-m}(k_0 x) = C_{m}\ H^{(1)}_m(k_0R_2),\\
\label{Eq:PositivePerfectCavityDispersionRelationEnd}
F_{-m}\ \lim_{x \to 0}\left[\frac{(-x)}{R_2}J'_{-m}(k_0 x)\right]+ G_{-m}\ \lim_{x \to 0}\left[\frac{(-x)}{R_2}Y'_{-m}(k_0 x)\right] \nonumber \\\qquad\qquad\qquad\qquad= C_{m}\ H'^{(1)}_m(k_0R_2).
\end{eqnarray}
Following the same argumentation as for equations (\ref{Eq:PerfCafIntr3}) and (\ref{Eq:PerfCafIntr4}), we find that equations (\ref{Eq:PositivePerfectCavityDispersionRelationMiddle1}) and (\ref{Eq:PositivePerfectCavityDispersionRelationEnd}) can be solved if $G_{-m} = 0$ and $C_m = 0$ for all azimuthal mode numbers $m\neq0$, without any restrictions on the values of $F_{-m}$ and $k_0$. 
We can reinsert this in the boundary conditions at $\rho = R_1$:
\begin{eqnarray}
\label{Eq:PositivePerfectCavityDispersionRelationBegin2}
A_m\ J_m(k_0R_1) = F_{-m}\ J_{-m}(k_0 R_1),\\
\label{Eq:PositivePerfectCavityDispersionRelationMiddle12}
A_m\ J'_m(k_0R_1) = F_{-m}\ (-1)J'_{-m}(k_0 R_1).
\end{eqnarray}
Using the identity $J_{-m}(x) = (-1)^mJ_m(x)$, it is clear that the set can be solved for all frequencies for which $J_m(k_0R_1) = 0$ or $J'_m(k_0R_1) = 0$. Depending on the angular mode number $m$, $A_m$ then equals $F_{-m}$ or $-F_{-m}$.

The eigenfrequencies of this cavity are defined by the zeros of the Bessel's function or its derivative: 
\begin{eqnarray}
&k_0 = j_{m,n}/R_1,\\
&k_0 = j'_{m,n}/R_1,
\end{eqnarray}
where $j_{m,n}$ and $j'_{m,n}$ are the $n$th solution of $J_m(x)=0$ and $J'_m(x)=0$ respectively.
The eigenfrequencies, therefore, cannot be chosen at will. To overcome this limitation we modify the design by replacing the vacuum in the inner region (I) by a transformation-optical material that maps the radial coordinate onto a larger one, i.e., $f(R_1) =f_{R_1}> R_1 $. Inside the inner region we implement the transformation given by $f(\rho) = f_{R_1}\rho/R_1$. Obviously, the transformation in region (II) should also map $R_1$ onto $f_{R_1}$. This modification allows us to design the cavity to have perfectly confined modes at arbitrary frequencies, since the resulting dispersion relation is given by:
\begin{eqnarray}
\label{Eq:PositivePerfectCavityDispersionRelationBegin3}
A_m\ J_m(k_0f_{R_1}) = F_{-m}\ J_{-m}(k_0 f_{R_1}),\\
\label{Eq:PositivePerfectCavityDispersionRelationMiddle13}
A_m\ J'_m(k_0f_{R_1}) = F_{-m}\ (-1)J'_{-m}(k_0 f_{R_1}).
\end{eqnarray}
This cavity has solutions for $k_0f_{R_1} = j'_{m,n}$ or $k_0f_{R_1} = j_{m,n}$. Equivalently, we can write
\begin{eqnarray}
\label{Eq:SolutionNewRightHandedCavity1}
&\frac{R_2}{\lambda_0} = \frac{R_2}{f_{R_1}}\frac{j_{m,n}}{2\pi},\\
\label{Eq:SolutionNewRightHandedCavity2}
&\frac{R_2}{\lambda_0} = \frac{R_2}{f_{R_1}}\frac{j'_{m,n}}{2\pi}.
\end{eqnarray}

\begin{figure}[tb]
  \centering
   \includegraphics[]{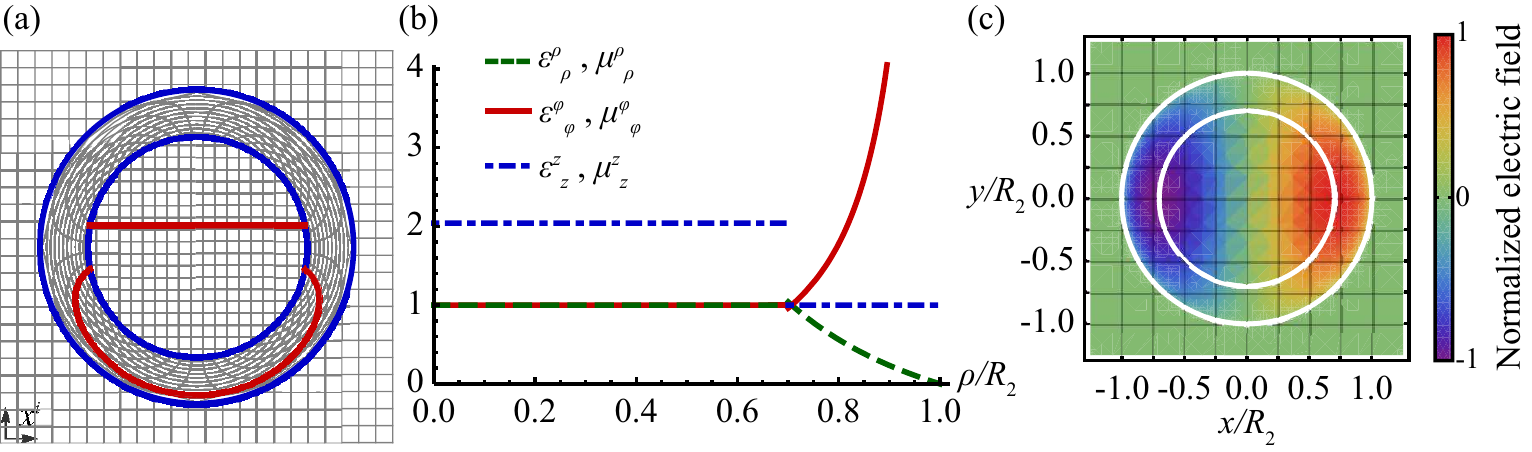} 
  \caption{(a) The grid lines of a right-handed cavity in which subwavelength modes can be confined in the absence of radiation losses. These grid lines correspond to a transformation defined by equations (\ref{Eq:SubwavelengthRightHandedCavityRegionI})-(\ref{Eq:SubwavelengthRightHandedCavityRegionII2}) (b) The material parameters required to materialize this coordinate transformation. These material parameters were considerably simplified by choosing $R_1 = 0.7R_2$  (c) The electric field mode profile of a perfectly confined mode inside the cavity ($m=1$, $\lambda_0 \approx 3.4 R_2$).}
  \label{Fig:RightHandedCavity_figures}
\end{figure}

In figure \ref{Fig:RightHandedCavity_figures}(a), we plot the grid lines of such a cavity in which the inner radius is mapped onto a larger value: $f_{R_1} = R_2$. The underlying coordinate transformations are
\begin{equation}
\label{Eq:SubwavelengthRightHandedCavityRegionI}
f(\rho) = \frac{R_2}{R_1}\rho,
\end{equation}
in region (I), and
\begin{eqnarray}
\label{Eq:SubwavelengthRightHandedCavityRegionII1}
f(\rho) = \frac{R_2}{\sqrt{R_1^2-R_2^2}}\sqrt{\rho^2-R_2^2},\\
\label{Eq:SubwavelengthRightHandedCavityRegionII2}
g(\phi) = -\phi,
\end{eqnarray}
in region (II).
The thick red line clearly indicates that the coordinates in the inner region are not continuously guided into region (II). The necessary condition for reflectionless transformation media is not valid since at the interface $\rho = R_1$ the coordinates of region (II) cannot be matched with those of region (I) through a combination of rotation and displacement  \cite{Qiu-2008}. This cavity, therefore, only confines light at discrete resonance frequencies. This gives a geometrical explanation of the discreteness of the solutions as given by equations (\ref{Eq:SolutionNewRightHandedCavity1})-(\ref{Eq:SolutionNewRightHandedCavity2}). Figure \ref{Fig:RightHandedCavity_figures}(b) shows the material parameters of this cavity. Three elements of the material tensors are simplified to the vacuum values thanks to the particular choice of $R_1 = 0.7R_2$. Finally, in figure \ref{Fig:RightHandedCavity_figures}(c), we plot the electric field of a perfectly confined mode inside this cavity. There are no fields outside the cavity and the cavity is subwavelength ($\lambda_0 = 3.4 R_2$).

\section{Conclusion}
In this paper, we have discussed designs of electromagnetic cavities based on transformation optics. We derived the dispersion relations of cavity structures based on radial and azimuthal coordinate transformations and applied those to calculate their bound modes.
Some of these transformations enlarge the optical path length inside the cavity, whereas others are based on a folding of the electromagnetic space. Finally, we have shown how the combination of radial and azimuthal transformations can eliminate the left-handedness of the perfect cavity, while preserving its most important characteristics: confinement of electromagnetic modes with unlimited quality factor due to radiation losses, even if the wavelength is larger than the dimensions of the cavity.

\ack
Work at the Vrije Universiteit Brussel was supported by BelSPO Grant No.\ IAP6/10 Photonics@be, the FWO-Vlaanderen, and the Research Council (OZR) of the VUB. Work at Ames Laboratory was supported by the U.S. Department of Energy, Office of Basic Energy Science, Division of Materials Sciences and Engineering (Ames Laboratory is operated for the U.S. Department of Energy by Iowa State University under Contract No. DE-AC02-07CH11358). V.~G. is a Research Assistant (Aspirant) of the Research Foundation-Flanders (FWO-Vlaanderen). P.~T. acknowledges the Belgian American Educational Foundation for a fellowship.

\newpage

\bibliographystyle{iopart-num}

\begin{thebibliography}{10}
\expandafter\ifx\csname url\endcsname\relax
  \def\url#1{{\tt #1}}\fi
\expandafter\ifx\csname urlprefix\endcsname\relax\def\urlprefix{URL }\fi
\providecommand{\eprint}[2][]{\url{#2}}

\bibitem{Walther-2006}
Walther H, Varcoe B~T~H, Englert B~G and Becker T {2006} {\em
  {Rep.~Prog.~Phys.}\/} {\bf {69}} {1325--1382}

\bibitem{Miller-2005}
Miller R, Northup T, Birnbaum K, Boca A, Boozer A and Kimble H {2005} {\em
  {\jpb}\/} {\bf {38}} {S551--S565}

\bibitem{Klaers-2010}
{Klaers} J, {Schmitt} J, {Vewinger} F and {Weitz} M 2010 {\em Nature\/} {\bf
  468} 545--548

\bibitem{Matsko-2009}
{Matsko} A~B 2009 {\em Practical Applications Of Microresonators In Optics And
  Photonics\/} (London: Taylor and Francis)

\bibitem{Vahala-2003}
{Vahala} K~J 2003 {\em Nature\/} {\bf 424} 839--846

\bibitem{Ilchenko-1996}
Gordetsky M~L, Savchenkov A~A and Ilchenko V~S 1996 {\em Opt.~Lett.\/} {\bf 21}
  453--455

\bibitem{Armani-2003}
{Armani} D~K, {Kippenberg} T~J, {Spillane} S~M and {Vahala} K~J 2003 {\em
  Nature\/} {\bf 421} 925--928

\bibitem{Hinds-1990}
Hinds E~A 1990 {\em Adv.~at.~molec.~opt.~Phys.\/} {\bf 28} 237--289

\bibitem{Kimble-1998}
Kimble H~J 1998 {\em Physica Scripta\/} {\bf T76} 127--137

\bibitem{Hinds-2007}
Trupke M, Metz J, Beige A and Hinds E~A 2007 {\em J. Mod. Optic.\/} {\bf 54}
  1639--1655

\bibitem{Chang-1996}
{Chang} R~K and {Campillo} A~J 1996 {\em Optical Processes In Microcavities
  Advanced Series in Applied Physics vol 3\/} (Singapore: World Scientific)

\bibitem{Kavokin-2006}
{Kavokin} A~V, {Baumberg} J~J, {Malpuech} G and {Laussy} F~P 2006 {\em
  Microcavities\/} (Oxford: Oxford University Press)

\bibitem{Vahala-2009}
Min B, Ostby E, Sorger V, Ulin-Avila E, Yang L, Zhang X and Vahala K {2009}
  {\em {Nature}\/} {\bf {457}} {455--459}

\bibitem{Leonhardt-2006}
{Leonhardt} U 2006 {\em Science\/} {\bf 312} 1777--1780

\bibitem{Pendry-2006}
{Pendry} J~B, {Schurig} D and {Smith} D~R 2006 {\em Science\/} {\bf 312}
  1780--1782

\bibitem{Leonhardt-2009}
{Leonhardt} U and {Philbin} T~G 2009 {\em Prog.\ Opt.\/} {\bf 53} 69--152

\bibitem{Balazs-1957}
{Balazs} N~L 1957 {\em Phys.\ Rev.\/} {\bf 110} 236--239

\bibitem{Plebanski-1960}
{Plebanski} J 1960 {\em Phys.\ Rev.\/} {\bf 118} 1396--1408

\bibitem{Felice-1971}
{Felice} D~F 1971 {\em Gen.\ Rel.\ Grav.\/} {\bf 2} 347--357

\bibitem{Ward-1996}
{Ward} A~J and {Pendry} J~B 1996 {\em J.\ Mod.\ Opt.\/} {\bf 43} 773--793

\bibitem{Chen-2010}
Chen H, Chan C~T and Sheng P {2010} {\em {Nature Mater.}\/} {\bf {9}}
  {387--396}

\bibitem{Shurig-2007}
{Shurig} D, {Pendry} J~B and {Smith} D~R 2007 {\em Opt.\ Express\/} {\bf 15}
  14772--14782

\bibitem{Tsang-2007}
{Tsang} M and {Psaltis} D 2007 {\em Phys.~Rev.~B.\/} {\bf 77} 35122

\bibitem{Yan-2008}
{Yan} M, {Yan} W and {Qiu} M 2008 {\em Phys.\ Rev.\ B\/} {\bf 78} 125113

\bibitem{Leonhardt-2010}
Leonhardt U and Philbin T~G {2010} {\em {Phys.~Rev.~A.}\/} {\bf {81}}
  {011804(R)}

\bibitem{Rahm3-2008}
{Rahm} M, {Cummer} S~A, {Schurig} D, {Pendry} J~B and {Smith} D~R 2008 {\em
  Phys.~Rev.~Lett.\/} {\bf 100} 63903

\bibitem{Kwon-2008}
{Kwon} D and {Werner} D~H 2008 {\em Opt.~Express\/} {\bf 16} 18731--18738

\bibitem{Wee-2009}
Wee W~H and Pendry J~B 2009 {\em New. J. Phys.\/} {\bf 11} 073033

\bibitem{Wei-2008}
{Yan} W, {Yan} M, {Ruan} Z and {Qiu} M 2008 {\em New\ J.\ Phys.\/} {\bf 10}
  043040

\bibitem{Leonhardt2-2009}
{Leonhardt} U and {Tyc} T 2009 {\em Science\/} {\bf 323} 110--112

\bibitem{Kildishev-2008}
Kildishev A~V, Cai W, Chettiar U~K and Shalaev V~M 2008 {\em New J. Phys.\/}
  {\bf 10} 115029

\bibitem{Greenleaf-2009}
{Greenleaf} A, {Kurylev} Y, {Lassas} M and {Uhlmann} G 2009 {\em SIAM Review\/}
  {\bf 51} 3--33

\bibitem{Chan-2009}
Lai Y, Chen H, Zhang Z~Q and Chan C~T 2009 {\em Phys. Rev. Lett.\/} {\bf 102}
  093901

\bibitem{Lai-2009}
Lai Y, Ng J, Chen H, Han D, Xiao J, Zhang Z~Q and Chan C~T 2009 {\em Phys. Rev.
  Lett.\/} {\bf 102} 253902

\bibitem{Leonhardt2-2006}
{Leonhardt} U and {Philbin} T~G 2006 {\em New {J.}\ Phys.\/} {\bf 8} 247

\bibitem{Ginis-2010}
{Ginis} V, {Tassin} P, {Craps} B and {Veretennicoff} I 2010 {\em Opt.
  Express\/} {\bf 18} 5350--5355

\bibitem{Cummer-2010}
Cummer S~A and Thomson R~T 2010 {\em J. Opt.\/} {\bf 13} 024007

\bibitem{Miao-2010}
Miao R~X, Zheng R and Li M 2011 {\em Phys.~Lett.~B\/} {\bf 696} 550--555

\bibitem{Philbin-2008}
Philbin T~G, Kuklewicz C, Robertson S, Hill S, K\"onig F and Leonhardt U 2008
  {\em Science\/} {\bf 319} 1367--1370

\bibitem{Rubino-2011}
Rubino E, Belgiorno F, Cacciatori S~L, Clerici M, Gorini V, Ortenzi G, Rizzi L,
  Sala V~G, Kolesik M and Faccio D 2011 {\em New J. Phys.\/} {\bf 13} 085005

\bibitem{Faccio-2011}
Faccio D 2012 {\em Cont.\ Phys.\/} {\bf 53} 97--112

\bibitem{Mackay-2011}
Mackay T~G and Lakhtakia A 2011 {\em Phys. Rev. B\/} {\bf 83} 195424

\bibitem{McCall-2011}
{McCall} M~W, {Favaro} A, {Kinsler} P and {Boardman} A 2011 {\em J.~Opt.\/}
  {\bf 13} 024003

\bibitem{Smith-2004}
Smith D~R, Pendry J~B and Wiltshire M~C~K 2004 {\em Science\/} {\bf 305}
  788--792

\bibitem{Veselago-1968}
Veselago V~G 1968 {\em Sov.\ Phys.\ Usp.\/} {\bf 10} 509--514

\bibitem{Pendry-2000}
{Pendry} J~B 2000 {\em Phys.\ Rev.\ Lett.\/} {\bf 85} 3966--3969

\bibitem{Engheta-2002}
Engheta N 2002 {\em IEEE Ant. Wireless Prop. Lett.\/} {\bf 1} 10--13

\bibitem{Alu-2007}
Alu A, Engheta N, Erentok A and Ziolkowski R~W 2007 {\em IEEE Trans. Antennas
  Propag.\/} {\bf 49} 23--36

\bibitem{Tassin-2008}
Tassin P, Sahyoun X and Veretennicoff I 2008 {\em Appl. Phys. Lett.\/} {\bf 92}
  203111

\bibitem{Soukoulis-2009}
Wang B, Zhou J, Koschny T, Kafesaki M and Soukoulis C~M 2009 {\em J. Opt. A:
  Pure Appl. Opt.\/} {\bf 11} 114003

\bibitem{Wegener-2009}
Gansel J~K, Thiel M, Rill M~S, Decker M, Bade K, Saile V, {von Freymann} G,
  Linden S and Wegener M 2009 {\em Science\/} {\bf 325} 1513--1515

\bibitem{Papasimakis-2008}
Papasimakis N, Fedotov V~A, Zheludev N~I and Prosvirnin S~L 2008 {\em Phys.
  Rev. Lett.\/} {\bf 101} 253903

\bibitem{Tassin-2009}
Tassin P, Zhang L, Koschny T, Economou E~N and Soukoulis C~M 2009 {\em Phys.
  Rev. Lett.\/} {\bf 102} 053901

\bibitem{Giessen-2009}
Liu N, Langguth L, Weiss T, Kastel J, Fleischhauer M, Pfau T and Giessen H 2009
  {\em Nature Mater.\/} {\bf 8} 758--762

\bibitem{Zheludev-2010}
Tanaka K, Plum E, Ou J~Y, Uchino T and Zheludev N~I 2010 {\em Phys. Rev.
  Lett.\/} {\bf 105} 227403

\bibitem{Zharova-2008}
{Zharova} N~A, {Shadrivov} I~V and {Kivshar} Y~S 2008 {\em Opt.\ Express\/}
  {\bf 16} 4615--4620

\bibitem{Zhai-2010}
Zhai T, Zhou Y, Shi J, Wang Z, Liu D and Zhou J 2010 {\em Opt. Express\/} {\bf
  18} 11891--11897

\bibitem{Ginis-2010-2}
{Ginis} V, {Tassin} P, {Soukoulis} C~M and {Veretennicoff} I 2010 {\em Phys.
  Rev. B\/} {\bf 82} 113102

\bibitem{Chan-2008}
Chen H, Luo X, Ma H and Chan C~T 2008 {\em Opt. Express\/} {\bf 16}
  14603--14608

\bibitem{Engheta-2009}
Castaldi G, Gallina I, Galdi V, Al\`{u} A and Engheta N 2009 {\em Opt.
  Express\/} {\bf 17} 3101--3114

\bibitem{Tassin-2006}
Tassin P, Veretennicoff I and {Van Der Sande} G 2006 {\em Opt. Commun.\/} {\bf
  264} 130--134

\bibitem{Qiu-2008}
Yan W, Yan M and Qiu M 2008 {\em arXiv:0806.3231v1\/}

\end{thebibliography}
\section*{References}
\providecommand{\newblock}{}

\end{document}